\documentclass[prc,twocolumn,epsfig,nofootinbib,floatfix]{revtex4}
\usepackage{graphics}
\usepackage{epsfig}
\usepackage{amsfonts}
\usepackage{amsmath}
\usepackage{bm}

\usepackage{color}

\begin{document}
\title{Deformation effects in toroidal and compression dipole
excitations of $^{170}$Yb: Skyrme-RPA analysis\\
}
\author{J. Kvasil$^{1}$,  V.O. Nesterenko  $^{2}$,
W. Kleinig  $^{2,3}$, and P.-G. Reinhard $^{4}$}
\affiliation{$^{1}$ \it Institute of Particle and Nuclear Physics,
Charles University, CZ-18000, Prague, Czech Republic}
{\email{kvasil@ipnp.troja.mff.cuni.cz}} \affiliation{$^{2}$   \it
Laboratory of Theoretical Physics, Joint Institute for Nuclear
Research, Dubna, Moscow region, 141980, Russia}
\email{nester@theor.jinr.ru} \affiliation{$^{3}$  \it Technische
Universit\"at Dresden,
  Institut f\"ur Analysis, D-01062, Dresden, Germany}
\affiliation{$^{4}$
  \it Institut f\"ur Theoretische Physik II,
     Universit\"at Erlangen, D-91058, Erlangen, Germany}


\begin{abstract}
The effect of nuclear deformation on the isoscalar toroidal and
compression dipole modes in prolate $^{170}$Yb is studied in the
framework of the random-phase-approximation method with a
representative set of Skyrme forces (SV-bas, SLy6, SkM$^*$ and
SkI3). It is shown that the deformation crucially redistributes
the strength of both modes. The compression mode has the same
sequence of $\mu$=0 and 1 branches as the isovector giant dipole
resonance where for prolate nuclei the $\mu=0$ mode is lower in
energy ($\mu$ being the projection of the axial momentum of the
mode). Instead, the toroidal mode exhibits an anomalous (opposite)
sequence where the $\mu$=1 branch precedes the $\mu$=0 one.
\end{abstract}

\maketitle

\section{Introduction}

The toroidal and compression modes (TM and CM in what follows) in
the isoscalar dipole E1(T=0) channel represent two unconventional
kinds dipole motion, attracting increasing interest last years
\cite{Pa07}. Both modes are second-order corrections to the
dominant giant dipole resonance (GDR) flow \cite{Kv11}. After
extraction of the spurious center-of-mass motion, these modes
become dominant in E1(T=0) channel. TM and CM were observed in $
(\alpha,\alpha')$ scattering experiments \cite{Youngb04,Uchi04}
where they were treated as the low-energy (TM) and high-energy
(CM) branches of the isoscalar dipole giant resonance (ISGDR)
\cite{Pa07}. The modes are strongly mixed \cite{Pa07,Kv11}.

The TM is viewed as a toroid-like {\it vortical} flow
\cite{Du75,Se81}. Its restoring force is caused by distortions of
the Fermi sphere in the momentum space. So the TM can be treated
as a specific transversal oscillation of an elastic globe
\cite{Bast93,Mis06}. As was recently shown \cite{Rep_PRC13}, the
TM lies at the low-energy region of what is called the pygmy
dipole resonance (PDR) and determines the basic collective flow in
the region. This is important since this low lying strength
carries a great deal of information on basic nuclear parameters as
incompressibility, symmetry energy, and effective masses
\cite{Rei13c} which, in turn, is related to questions of the
neutron skin and neutron equation-of-state. The CM  exhibits an
{\it irrotational} compression dipole flow \cite{Ha77,St82}.
Actually, the TM and CM are transversal (vortical) and
longitudinal (irrotational) counterparts.

The TM and CM were thoroughly investigated in various models, see
extensive references in \cite{Pa07,Kv11,Kv03,Kva_EPJA13}. Most
studies so far were limited to spherical nuclei. At the same time,
effects of nuclear deformation in TM and CM can be very strong. As
shown in our recent study for the chain of Sm isotopes
\cite{Kva_EPJA13}, the deformation appreciably reshuffles the
TM/CM strengths. In particular, there was a spectacular
deformation splitting (exceeding 5 MeV) of CM and considerable
downshift of the TM strength (which can affect the PDR
properties).

In this paper, we continue investigation of the deformation
effects in TM and CM, now for a typical rare-earth axial nucleus
$^{170}$Yb which exhibits a sizable prolate deformation. Like in
the previous study \cite{Kva_EPJA13}, we employ the separable
random-phase-approximation (SRPA) approach using the Skyrme
energy-density functional \cite{Ne02,Ne06}. SRPA was already
successfully used in various studies in spherical and deformed
nuclei (electric
\cite{Kv11,Kva_EPJA13,nest_IJMPE_07_08,kle_PRC_08,Kva_IJMPE_12}
and magnetic  \cite{Ve09,Nest_JPG_10,Nest_IJMPE_10} giant
resonances, E1 strength near the particle thresholds
\cite{Kva_IJMPE_09,Kva_IJMPE_11}). Here we employ it with the
Skyrme forces covering various values of  isoscalar effective mass
$m^*_0/m$. These are SV-bas \cite{Kl09}, SkM*\cite{Ba82}, SLy6
\cite{Ch97}, SkI3 \cite{Re95} with $m^*_0/m$=0.90, 0.79, 0.69,
0.58, respectively. While in our previous study \cite{Kva_EPJA13}
the CM deformation effects were mainly analyzed, here we consider
deformation features of both TM and CM. It will be shown that,
unlike the isovector giant dipole resonance (GDR) and CM, the TM
demonstrates an anomalous deformation splitting. Namely, in TM the
dipole branch $\mu$=1 lies lower than the $\mu$=0 one, in contrast
to the opposite sequence 0/1 typical for prolate quadrupole
shapes.

The paper is organized as follows. In Sec. 2 the calculation
scheme is outlined. In Sec. 3 the main results are discussed. In
Sec. 4, the conclusions are given.

\section{Calculation scheme}

The study is performed within the SRPA approach with
the Skyrme functional
\cite{Ne02,Ne06}. The method is fully self-consistent as i) both
the mean field and residual interaction are obtained from the same
Skyrme functional \cite{Skyrme,Vau72,Ben03}, ii)  the residual
interaction includes all the functional contributions as well as
the Coulomb (direct and exchange) terms. The self-consistent
factorization of the residual interaction in SRPA dramatically
reduces the computational effort for deformed nuclei while keeping
high accuracy of the method. As shown in the systematic SRPA study
of GDR in rare-earth and actinide regions \cite{kle_PRC_08}, the
method provides an excellent description of the experimental data
in a wide manifold of deformed nuclei.

In order to test the sensitivity of the results, four different
Skyrme forces (SV-bas \cite{Kl09}, SkM*\cite{Ba82}, SLy6
\cite{Ch97}, and SkI3 \cite{Re95}) with various isoscalar
effective masses are used. The 2D representation in cylindrical
coordinates with a mesh size of 0.7 fm and a calculation box of 21
fm is exploited. The pairing (with delta forces) is treated at the
BCS level \cite{Ben00}. The equilibrium axial quadrupole
deformation is determined by minimization of the total energy. In
$^{170}$Yb, we obtain for all four Skyrme forces the deformation
parameter $\beta_2$=0.34-0.35 and corresponding quadrupole moment
$Q_2\approx$ 8.5 b. These values are in acceptable agreement with
the experimental data $\beta^{\mathrm{exp}}_2$=0.32 and
$Q^{\mathrm{exp}}_2$=7.6 b \cite{Raman87}.

The TM and CM strength functions read
\begin{equation} \label{39}
 S\:_{\gamma}(E1\mu,\;E) =  \sum_{\nu} \:
|\:\langle\nu|\:\hat{M}_{\gamma}(\;E1\mu)\:|0\rangle \:|^2
\:\xi_{\Delta}(E-E_{\nu})
\end{equation}
where $\xi_{\Delta}(E-E_{\nu}) = \Delta/[2\pi((E-E_{\nu})^2 +
(\Delta/2)^2] $ is the Lorentz weight with the averaging parameter
$\Delta$= 1 MeV. Further, $|0\rangle$ is the ground state wave
function, $E_{\nu}$ and $|\nu\rangle$ are the energy and wave
function of the $\nu$-th RPA state. $\hat{M}_{\gamma}(\;E1\mu)$ is
the transition dipole operator where $\gamma$ labels the cases TM,
CM, or GDR. The TM and CM operators read \cite{Kv11}:
\begin{eqnarray} \label{29}
&& \hat{M}_{\rm{tor}} (E1\mu) = - \frac{2}{2c\sqrt{3}}\:\int\:d^3r
\: \hat{\vec{j}}_{\rm{nuc}}(\vec{r})
\nonumber \\
&&  \cdot\:\left[\:\frac{\sqrt{2}}{5} \:r^2 \:\vec{Y}_{12\mu} +
(r^2 - \delta_{T,0} \langle
r^2\rangle_0)\:\vec{Y}_{10\mu}\:\right],
\end{eqnarray}
\begin{equation} \label{30}
\hat{M}_{\rm{com}} (E1\mu) = \frac{1}{10} \int\:d^3r \:
\hat{\rho}(\vec{r})
\left[r^3 - \delta_{T,0}\:\frac{5r}{3} \langle r^2 \rangle_0
\right] Y_{1\mu},
\end{equation}
where $\hat{\rho}(\vec{r})$ and
$\hat{\vec{j}}_{\rm{nuc}}(\vec{r})$ are operators of nuclear
density and  convection current, respectively. Symbols $Y_{\lambda
\mu}$ and $\vec{Y}_{\lambda l \mu}$ stand for spherical harmonics
and vector spherical harmonics. The value $\langle r^2\rangle_0 =
\int\:d^3r \: \rho_0(\vec{r}) \:r^2$ is the ground state
root-mean-square radius. The last terms in (\ref{29}) and
(\ref{30}) represent the center-of-mass corrections (c.m.c.)
active in T=0  channel \cite{Kv11}. The neutron and proton
effective charges  in this channel are
$e_n^{\rm{eff}}=e_p^{\rm{eff}}$=1.

The photo-absorption cross-section is determined through the
strength function (\ref{39}) as \cite{Ri80}:
\begin{equation} \label{43}
\sigma_{\rm{phot}}(E1\mu) = \frac{16\:\pi^3\:\alpha_{e}}{9\:e^2}
\;E \cdot S_{\rm{GDR}}(E1\mu,\:E)
\end{equation}
where $\alpha_{e}=1/137$ is the fine-structure constant and
\begin{equation}\label{E1}
 \hat{M}_{\rm{GDR}} (E1\mu)
  = \frac{N}{A}\sum_{p=1}^Z r_p Y_{1\mu}(\hat{r}_p)
  -
  \frac{Z}{A}\sum_{n=1}^N r_n Y_{1\mu}(\hat{r}_n)
\end{equation}
is the standard dipole isovector (T=1) transition operator.
\begin{figure}[t]
\includegraphics[width=10.0cm]{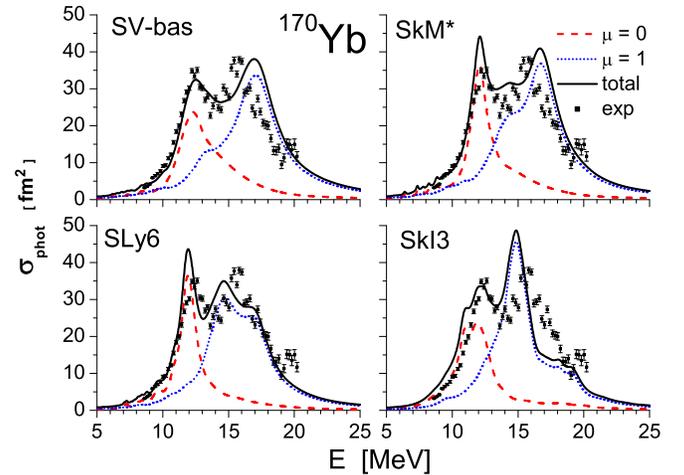}
\caption{\label{fig1} (color online) The SRPA total (black solid)
and partial $\mu$=0 (red dash) and $\mu$=1 (blue dotted)
photo-absorption cross sections in $^{170}$Yb, computed  for
various Skyrme forces as indicated. The experimental data
\protect\cite{Yb_exp1,Yb_exp2} are depicted by black full
squares.}
\end{figure}

The total strengths (\ref{39}) and (\ref{43}) are computed as the
sums of $\mu$=0 and twice the $\mu$=1 contributions. The input
SRPA operators are chosen following the prescription \cite{Kv11}.

The configuration space for the SRPA calculations covers all the
two-quasiparticle dipole states with the energy up to
$E_{\rm{cut}} \approx$ 175 MeV. The basis is sufficiently large to
exhaust nearly 100\% of the energy-weighted sum rules (EWSR): i)
Thomas-Reiche-Kuhn EWSR(T=1) for GDR \cite{Ri80} and ii) Harakeh's
EWSR(T=0) for CM \cite{Ha01}.  We need also this large basis to
lower the energy of the spurious E1(T=0) peak (c.m. mode) towards
its zero value. Here the spurious peak comes down to $1.5-3.0$ MeV
(depending on the Skyrme force) for CM and close to zero for TM.
This is safely below the regions of the studied TM and CM
strengths. Note that TM, driven mainly by vortical flow, is
generally less polluted by the spurious motion than the
irrotational CM. An extensive discussion of the sum rules and c.m.
correction in SRPA calculations can be found in Ref.
\cite{Kva_EPJA13}. Other details of the calculation scheme are
given elsewhere \cite{Kv11,Ne06,kle_PRC_08}.
\begin{figure}[t] 
\includegraphics[width=7.5cm]{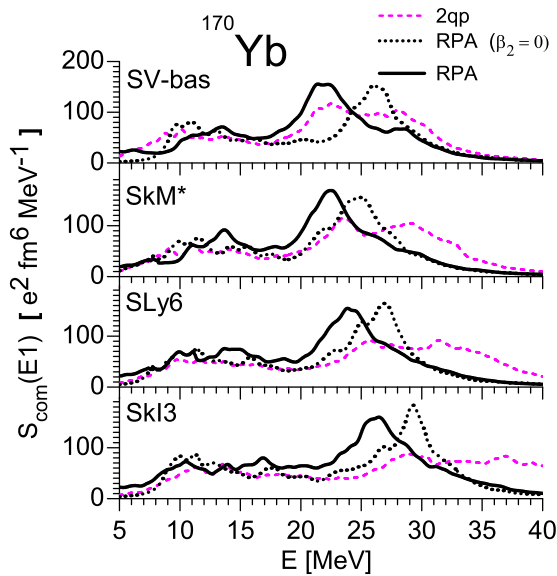}
\caption{\label{fig2}
 (color online) Compression  E1(T=0) strength functions in
$^{170}$Yb, calculated with different Skyrme parametrizations. The
2qp (red dash), RPA (black solid) and constrained (by the
spherical nuclear shape with $\beta_2$=0) RPA  (blue dotted)
strengths are depicted.}
\end{figure}
\begin{figure}[t] 
\includegraphics[width=7.5cm]{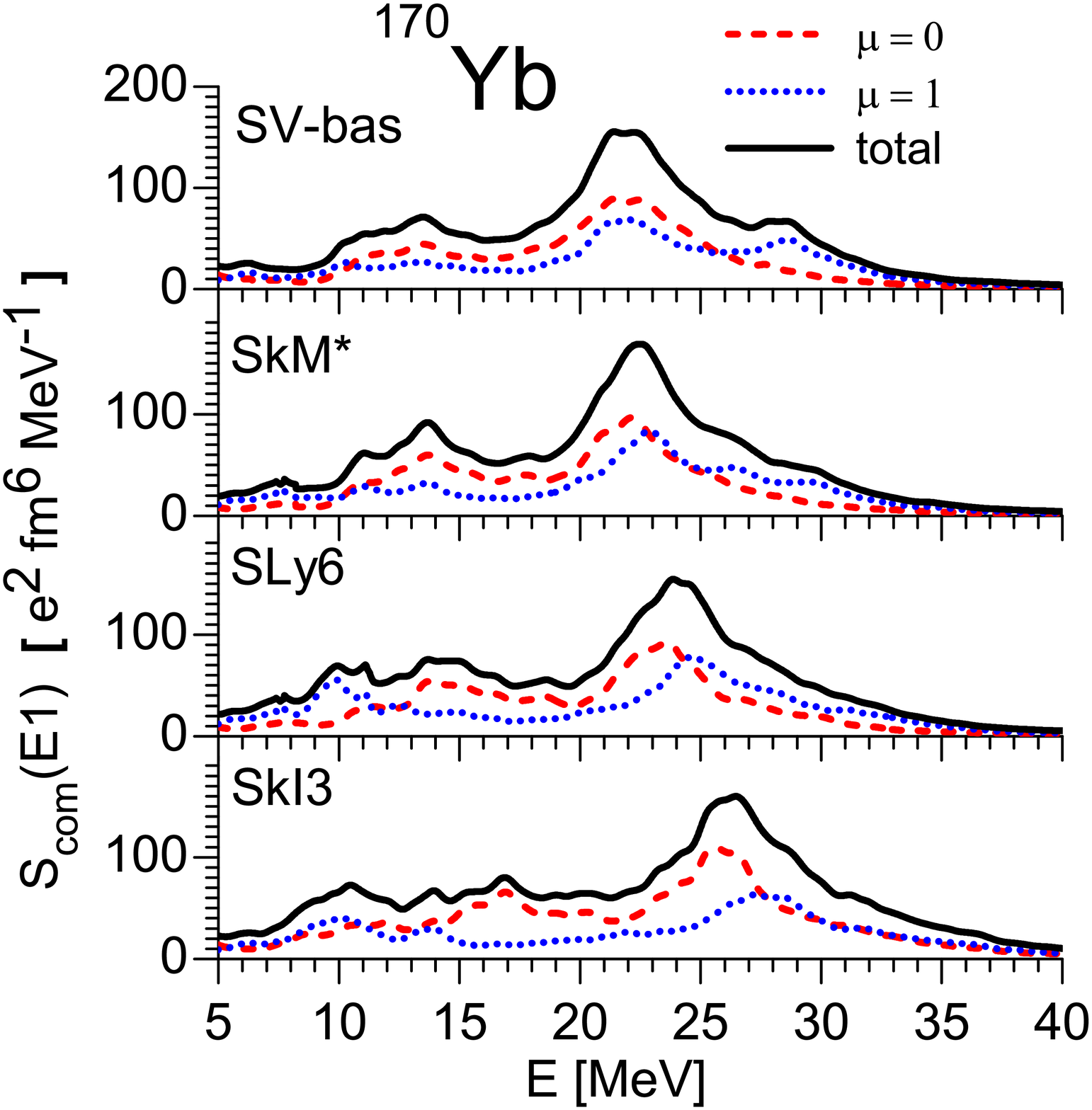}
\caption{\label{fig3} (color online) The total (solid black),
$\mu$=0 (red dash), and $\mu$=1 (blue dotted) RPA E1(T=0)
compression strength functions in $^{170}$Yb, calculated with
different Skyrme parametrizations.}
\end{figure}
\begin{figure}  
\includegraphics[width=7.5cm]{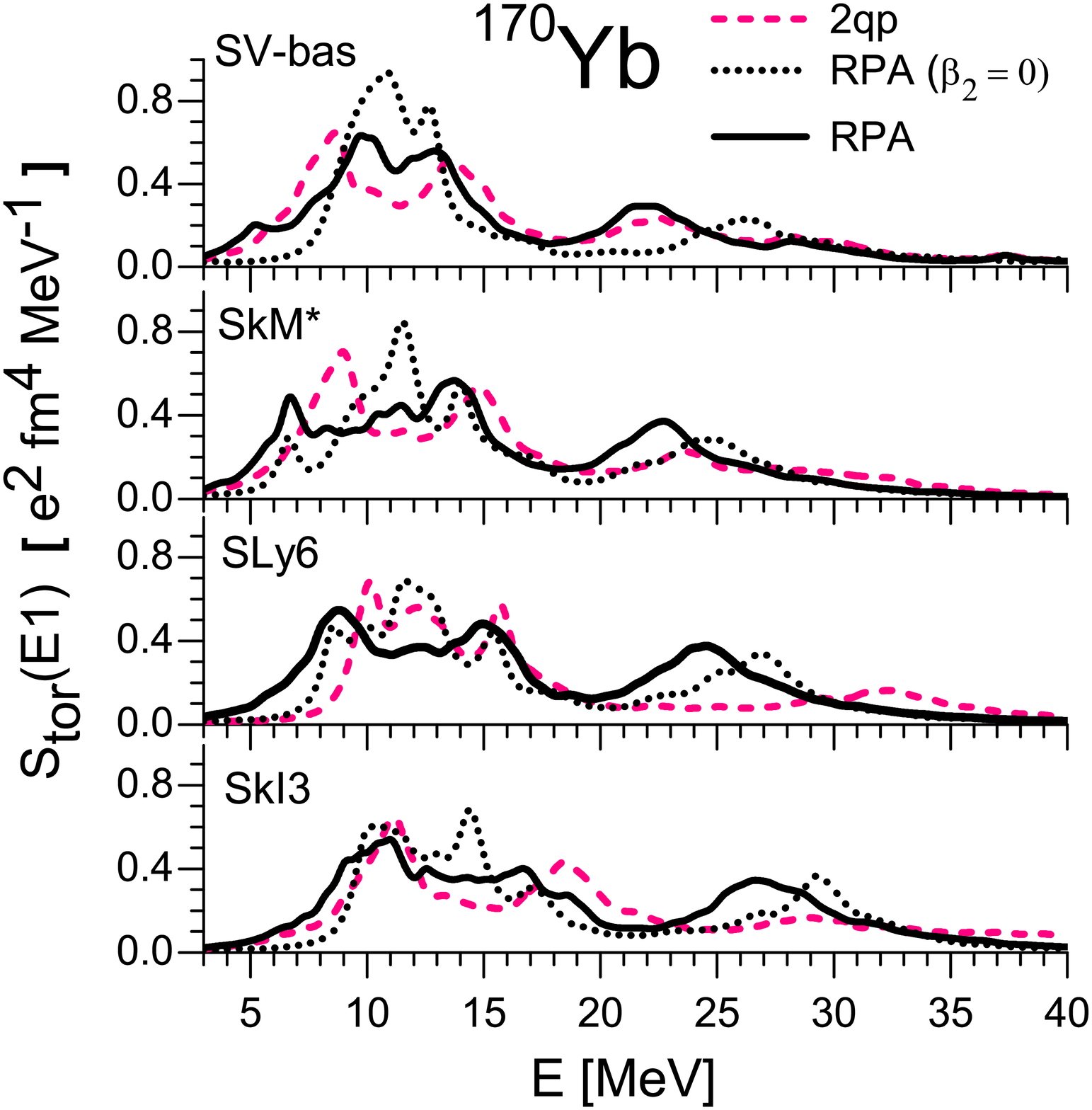}
\caption{\label{fig4} (color online) The same as in Fig. 2 but for
the toroidal E1(T=0) strengths.}
\end{figure}

\section{Results and discussion}

In Figure 1, the photo-absorption cross section calculated with
different Skyrme forces is compared to the experimental data
\cite{Yb_exp1,Yb_exp2}. Decomposition of the dipole strengths into
$\mu$=0 and $\mu$=1 branches is also given to demonstrate the
deformation splitting. The figure shows a good agreement with the
experiment for all the Skyrme forces. The energy position, width
and splitting of the GDR are well reproduced. At the same time,
the dependence of the description on the Skyrme force is also
visible.  A minor narrowing the GDR and downshift of $\mu$=1
branch from SV-bas to SkI3 (i.e. with decreasing the effective
mass $m^*_0/m$) are seen. They can be understood from the fact
that the level density of the single-particle spectra decreases
with decreasing $m^*_0/m$ \cite{Nest_PRC_04}. SV-bas somewhat
overestimates the GDR width and $\mu$=1 energy while SkI3
underestimates these values. The best result is obtained for SLy6
with $m^*_0/m$=0.69, which is in accordance to our previous
findings for the GDR within SRPA \cite{kle_PRC_08}. The
deformation splitting shows the standard ordering for the GDR in
prolate nuclei where the peak for $\mu=0$ lies energetically lower
than the $\mu$=1 peak.
\begin{figure} 
\includegraphics[width=7cm]{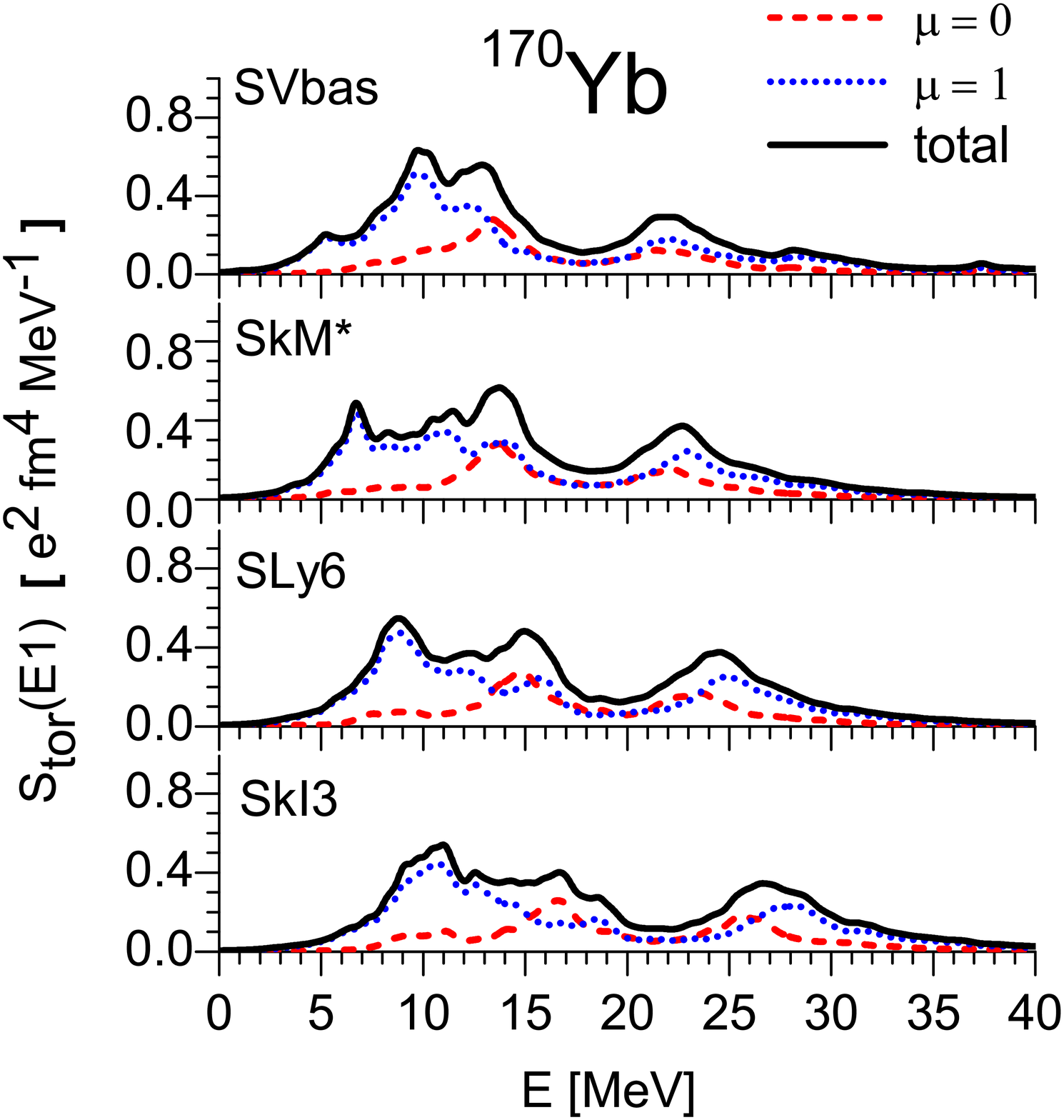}
\caption{\label{fig5} (color online) The same as in Fig. 3 but for
the toroidal E1(T=0) RPA strengths.}
\end{figure}

\begin{figure} 
\includegraphics[width=7cm]{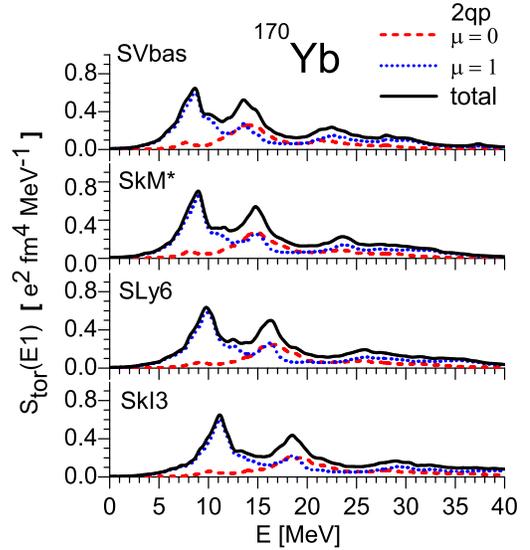}
\caption{\label{fig6} (color online) The same as in Fig. 3 but for
the toroidal E1(T=0) 2qp strengths.}
\end{figure}

In Figure 2, the CM E1(T=0) strength is presented.  The RPA residual
interaction significantly downshifts the strength as compared with the
unperturbed two-quasiparticle (2qp) spectra, which indicates a
considerable collectivity of the excitations. Like in Fig. 1 for GDR,
there is a visible regular dependence of the energy of the main CM
peak on the Skyrme force, which is explained by the compression of the
single-particle spectrum with increasing $m^*_0/m$. Fig. 2 shows a
strong impact of deformation. As compared with the spherical case (RPA
on a ground state with constraint on $\beta_2$=0), the deformation
downshifts the low-energy strength to the region 5-8 MeV. Besides, the
main CM peak is shifted from 26-30 MeV to 22-26 MeV. Similar
effects were earlier found for CM in Sm isotopes \cite{Kva_EPJA13}.

More information on the impact of deformation on the CM can be obtained
from Fig. 3, where the total RPA strength is given together with
its $\mu$=0 and 1 branches.  It is seen that the deformation
drastically redistributes the strength, making the ratio between
$\mu$=0 and 1 strengths dependent on the energy (while in the
spherical case this ratio is constant). In particular, we get the
dominance of $\mu$=0 strength at the central energy region 10-25
MeV and of $\mu$=1 strength at the peripheral regions at lower and
higher energies. If one considers the major and minor CM peaks at 22-26
and 26-32 MeV as $\mu$=0 and $\mu$=1 branches, then one may state
a considerable (though vague) deformation splitting. The
splitting changes from $\sim$7 MeV in SV-bas to $\sim$2 MeV in
SkI3. Like in the GDR, the deformation splitting exhibits the 0/1
order where the $\mu$=0 bump precedes the $\mu$=1 one.

Figures 4 and 5 illustrate the strength functions for TM. As compared
to high-energy CM, the TM strength is mainly concentrated at lower
energy 5-20 MeV. There is also a minor TM high-energy bump at
20-30 MeV, perhaps because of the CM and TM coupling. In Fig. 4,
we see the similar effects as in Fig. 2 for CM: i) downshift of
the RPA strength as compared to 2qp case, ii) strong
redistribution of the RPA strength due to deformation
(including a shift of the strength
to the low-energy region below 7-8 MeV and also a downshift of the minor
high-energy bump), and iii) visible dependence of the TM centroid
on $m^*_0/m$.

In Figure 5, we see a remarkable strong effect of deformation on
the low-energy TM: the deformation splitting results in an
overwhelming dominance of $\mu$=1 branch in the TM strength below
12-15 MeV. This is in contrast to the GDR where, following Fig. 1,
the $\mu$=0 branch dominates at $E <$ 13 MeV. Moreover,  TM has
the opposite order of the branches as compared to the GDR and CM
($\mu=1$ is lower than $\mu=0$). The difference takes place in
spite of the fact that both, TM and GDR, originate from the same
2qp dipole transitions with $\Delta \mathbb{N}$=1, where
$\mathbb{N}$ is the principle shell quantum number. As seen from
Fig. 6, the effect appears already for the unperturbed 2qp
strength and therefore has its origin in the mean field and not in
the recoupling by RPA. Perhaps it is caused by a different
character of the flows in TM and GR (vortical and irrotational,
respectively) and thus different 2qp excitations generating TM and
GDR. This effect calls for a careful microscopical analysis which
is the next step on our road map.

\section{Conclusions}

The effects of nuclear axial quadrupole deformation on the isoscalar
dipole compression and toroidal modes (CM and TM) were investigated in
prolate $^{170}$Yb within the random-phase approximation using the
Skyrme energy-density functional. The Skyrme parametrization SV-bas
\cite{Kl09}, SkM*\cite{Ba82}, SLy6 \cite{Ch97}, and SkI3 \cite{Re95}
were used which have different isoscalar effective mass $m^*_0/m$
varying from 0.9 for SV-bas to 0.58 for SkI3. All four forces provide
similar general pattern, but visible dependence on $m^*_0/m$ in the
details of the spectral distributions.

The strong impact of deformation found here is in accordance with our
previous study for Sm isotopes \cite{Kva_EPJA13}. The deformation
significantly redistributes the strengths, in particular the
ratios between $\mu$=0 and 1 branches. The $\mu$=0 strength
becomes dominant in a wide central region while the $\mu$=1 strength
takes the lead at lower and higher energy.

The most interesting deformation effect takes place for the low-energy
TM where an anomalous sequence (as compared to the giant dipole
resonance and CM) of the $\mu$=0 and 1 branches is found. Namely, in
contrast to the common order in prolate nuclei, for the TM the $\mu$=1
branch comes energetically lower than the $\mu$=0 one. This leads to a
strong dominance of $\mu$=1 over $\mu$=0 at the energy $E <$ 12-15 MeV
(often associated with the region of the pygmy dipole resonance).  This
feature was also found in other prolate nuclei (not shown in the
paper). Perhaps it is caused by a different character of the
modes,
vortical flow for the TM and irrotational flow for the GDR, and thus
different isoscalar two-quasiparticle
excitations generating these flows.

\section*{Acknowledgments}
The work was partly supported by the DFG RE 322/14-1, a GSI F+E,
Heisenberg-Landau (Germany - BLTP JINR), and Votruba - Blokhintsev
(Czech Republic - BLTP JINR) grants. W.K. and P.-G.R. are grateful
for the BMBF support under contracts 05P12ODDUE. The support of
the research plan MSM 0021620859 (Ministry of Education of the
Czech Republic) and the Czech Science Foundation project
P203-13-07117S are also appreciated.

\end{document}